\documentstyle{article}
\title{\sc The Quantum Sphaleron}
\author{ Pedro F. Gonz\'alez-D\'{\i}az.\\
Instituto de Matem\'aticas y F\'{\i}sica Fundamental\\
Consejo Superior de Investigaciones Cient\'{\i}ficas\\
Serrano 121, 28006 Madrid (SPAIN)\\
}
\date{March 1, 1993}
\begin{document}
\maketitle
\large
\setlength{\baselineskip}{0.5cm}

\vspace{2.5cm}

A gravitational instanton is found that can tunnel into a new
more stable vacuum phase where diffeomorphism invariance is
broken and pitchfork bifurcations develop.  This tunnelling
process involves a double sphaleron-like transition which is
associated with an extra level of quantization which is above
that is contained in quantum field theory.

\pagebreak

1. Many field theories have a degenerate vacuum structure
showing more than one potential minimum. Such a complicate
vacuum structure makes it possible the occurrence at zero energy
of quantum transitions describable as instantons between states
lying in the vecinity of different vacua. Instantons are
localized objects which correspond to solutions of the Euclidean
equations of motion with finite action [1,2]. On the other hand,
for nonzero energies, transitions between distinct vacua may
also occur classically by means of sphalerons. Whereas
instantons tunnel from one potential minimum to another by going
below the barrier, sphalerons do the transit over the barrier.
Sphalerons correspond to the top of this barrier and are
unstable classical solutions to the field equations which are
static and localized in space as well [3,4].

In this letter, we want to explore the possibility of new
sphaleron-like transitions which are classically forbidden
though they may still take place in the quantum-mechanical
realm. Such transitions would typically pertain to nonlinear
systems showing bifurcations phenomena.  In order to see how
these nonclassical transitions may appear, let us consider a
theory with the Euclidean action
\begin{equation}
S_{E}(\varphi , x)
=-\frac{1}{2}\int_{\eta_{i}}^{\eta_{f}}d\eta\varphi^{2}((\frac{dx}{d\eta})^{2}
+x^{2}-\frac{1}{2}m^{2}\varphi^{2}x^{4}),
\end{equation}
where $m$ is the generally nonzero mass of a dimensionless
constant scalar field $\varphi$ and $\eta$ denotes a
dimensionless Euclidean time $\eta=\int\frac{d\tau}{x}$.  The
sign for the Euclidean action would be positive when we choose
the usual Wick rotation $t\rightarrow -i\tau$ (clockwise) or
negative for a Wick rotation $t\rightarrow +i\tau$
(anti-clockwise). If the 'energy' of the particles is positive,
then one should use clockwise rotation, but if it is negative
the rotation would be anti-clockwise [5].  It will be seen later
on that (1) corresponds to the case of a massive field
conformally coupled to gravity, with $x$ playing the role of the
Robertson-Walker scale factor. Thus, since the gravitational
energy associated with the scale factor is negative, one should
rotate $t$ not to $-i\tau$, but to $+i\tau$, such as it is done
in (1), and therefore the semiclassical path integral involving
action $S_{E}$ may be generally interpreted as a probability for
quantum tunnelling [5].  In the classical case, if the field
$\varphi$ is axionic, then it becomes pure imaginary, i.e.
$\varphi=i\varphi_{0}$, with $\varphi_{0}$ real. In this case,
the potential becomes
\begin{equation}
V(x) =
\varphi_{0}^{2}(\frac{1}{2}x^{2}+\frac{1}{4}m^{2}\varphi_{0}^{2}x^{4}),
\end{equation}
the theory has just one zero-energy vacuum, and the solution to
the classical equations of motion is $\bar{x}=0$, with Euclidean
action $S_{E}=0$.

Expanding about the classical solution we obtain the usual path
integral at the semiclassical limit [6]
\begin{equation}
<0\mid e^{-HT/\hbar}\mid
0>=N[det(-\partial_{\tau}^{2}+\omega^{2})]^{-\frac{1}{2}}(1+O(\hbar)),
\end{equation}
where $N$ is a normalization constant and $\omega^{2}=V''(0)$,
with $'$ denoting differentiation with respect to $x$. The
ground-state solution to the wave equation for the operator
$-\partial_{\tau}^{2}+\omega^{2}$ which corresponds here to a
harmonic oscillator with substracted zero-point energy can be
written as
\[\psi_{0}=\omega^{-1}e^{-\omega T}\sinh[\omega(\tau+\frac{T}{2})]. \]
Then, the path integral becomes constant and given by
\begin{equation}
<0\mid e^{-HT/\hbar}\mid
0>=(\frac{\omega}{\pi\hbar})^{\frac{1}{2}}(1+O(\hbar)).
\end{equation}
If the zero-point energy had not been substracted, then Eqn. (4)
would also contain the well-known time-dependent factor
$e^{-\frac{1}{2}\omega T}$ [6].

Let us now consider $\varphi^{2}$ as being the control parameter
for the nonlinear dynamic problem posed by action (1). All the
values of $\varphi^{2}$ corresponding to an axionic classical
field will be negative and, in the classical case, can be
continuously varied first to zero (a zero potential critical
point) and then to positive values (the field $\varphi$ has
become real, no longer axionic). The associated variation of the
dynamics will represent a typical classical bifurcation process
that can finally lead to spontaneous breakdown of a given
symmetry [7]. In the semiclassical theory this generally is no
longer possible however. Not all values of the squared field
$\varphi^{2}$ are then equally probable. For most cases, large
ranges of $\varphi^{2}$-values along the bifurcation itinerary
are strongly suppressed, and hence the bifurcation mechanism
would not take place. Nevertheless, there could still be sudden
reversible quantum jumps from the most probable negative values
to the most probable positive values of $\varphi^{2}$. Such
jumps would be expressible as analytic continuations in the
field $\varphi$ to and from its real axis, or alternatively, in
$x$ to and from its imaginary values, lasting a very short time.
The system will first go from the bottom of potential (2) for
$\varphi^{2}<0$ to the sphaleron point of potential (see Fig.1)
\begin{equation}
V(x)=\varphi
'_{0}^{2}(-\frac{1}{2}x^{2}+\frac{1}{4}m^{2}\varphi_{0}'^{2}x^{4}),
\end{equation}
without changing position or energy, and then will be perturbed
about the sphaleron saddle point to fall into the broken vacua
where the broken phase condenses for a short while, to finally
redo all the way back to end up at the bottom of potential (2)
for $\varphi^{2}<0$ again. The whole process may be denoted as a
quantum sphaleron transition and it is assumed to last a very
short time and to occur at a very low frequency along the large
time $T$. Therefore, one can use a dilute spahaleron
approximation which is compatible with our semiclassical
approach.  Thus, for large $T$, besides individual quantum
sphalerons, there would be also approximate solutions consisting
of strings of widely separated quantum sphalerons. In analogy
with the instanton case [6], we shall evaluate the functional
integral by summing over all such configurations, with $n$
quantum sphalerons centered at Euclidean times
$\tau_{1}$,$\tau_{2}$,...,$\tau_{n}$.  If it were not for the
small intervals containing the quantum sphalerons, $V''$ would
equal $\omega^{2}$ over the entire time axis, and hence we would
obtain the same result as in (4). However, the small intervals
with the sphalerons correct this expression. In the dilute
sphaleron approximation, instead of (4), one has
\begin{equation}
(\frac{\omega}{\pi\hbar})^{\frac{1}{2}}(-\frac{1}{2}K_{sph})^{n}(1+O(\hbar)),
\end{equation}
where $K_{sph}$ is an elementary frequency associated to each
quantum sphaleron, and the sign minus accounts for the feature
that particles acquire a negative energy below the sphaleron
barrier (Fig.1). Note that the action changes sign as one goes
from potential (2) to potential (5) below such a barrier.  The
factor $\frac{1}{2}$ has been introduced to account for the
feature that the sphaleron transition must take two particles,
both at the same time, from the bottom of the potential (2) for
$\varphi^{2}<0$ to make the transition to the two minima of
potential (5) and back to the bottom of potential (2)
simultaneously. After integrating over the locations of the
sphaleron centers, the sum over $n$ sphalerons produces a path
integral
\begin{equation}
<0\mid e^{-HT/\hbar}\mid
0>_{sph}=(\frac{\omega}{\pi\hbar})^{\frac{1}{2}}
e^{-\frac{1}{2}K_{sph}T}(1+O(\hbar)),
\end{equation}
which is proportional to the semiclassical probability of
quantum tunnelling from $\bar{x}=0$ first to
$\bar{x}_{\pm}=\pm(m\varphi'_{0})^{-1}$ and then to $\bar{x}=0$
again.

In Eqn.(7) we have summed over any number of sphalerons, since
all the small time intervals start and finish on the axis $x=0$,
at the bottom of potential (2).  Approximation (7) corresponds
[6] to a ground-state energy $E_{0}=\frac{1}{2}\hbar K_{sph}$.
Thus, the effect of the quantum sphalerons should be the
creation of an extra nonvanishing zero-point energy which must
correspond to a further level of quantization which is over and
above that is associated with the usual second quantization of
the harmonic oscillator.

2. As pointed out before, an Euclidean action with the same form
as (1) arises in a theory where a scalar field $\Phi$ with mass
$m$ couples conformally to Hilbert-Einstein gravity. Restricting
to a Robertson-Walker metric with scale factor $a$ and Wick
rotating anti-clockwise, the Euclidean action for this case
becomes
\begin{equation}
I=-\frac{1}{2}\int d\eta
N(\frac{\dot{\chi}^{2}}{N^{2}}+\chi^{2}
-\frac{\dot{a}^{2}}{N^{2}}-a^{2}+m^{2}a^{2}\chi^{2}),
\end{equation}
where the overhead dot means differentiation with respect to the
conformal time $\eta=\int d\tau/a$,
$\chi=(2\pi^{2}\sigma^{2})^{\frac{1}{2}}a\Phi$, $N$ is the lapse
function and $\sigma^{2}=2G/3\pi$. The equations of motion
derived from (8) are (in the gauge $N=1$)
$\ddot{\chi}=\chi+m^{2}a^{2}\chi$ and
$\ddot{a}=a-m^{2}\chi^{2}a$. We note that these two equations
transform into each other by using the ansatz $\chi=ia$.
Invariance under such a symmetry manifests not only in the
equations of motion, but also in the Hamiltonian action and
four-momentum constraints. The need for Wick rotating
anti-clockwise becomes now unambiguous [8], since all the
variable terms in the action become associated with energy
contributions which are negative if symmetry $\chi=ia$ holds.
Without loss of generality, the equations of motion can then be
written as the two formally independent expressions
$\ddot{\chi}=\chi-m^{2}\chi^{3}$ and $\ddot{a}=a+m^{2}a^{3}$.
If $\chi=ia$, then $\Phi$ becomes a constant axionic field
$\Phi=i(2\pi^{2}\sigma^{2})^{-\frac{1}{2}}$.  Therefore,
re-expressing action $I$ in terms of the field $\chi$ alone, one
can write for the Lagrangian in the gauge $N=1$
\begin{equation}
L(\varphi,a)=-(\frac{1}{2}\varphi^{2}\dot{a}^{2}+\frac{1}{2}\varphi^{2}a^{2}
-\frac{1}{4}m^{2}\varphi^{4}a^{4}+\frac{1}{2}R_{0}^{2}),
\end{equation}
where $\varphi=\frac{\Phi}{m_{p}}$, $m_{p}$ is the Planck mass,
and $R_{0}^{2}$ is an integration constant which has been
introduced to account for the axionic character of the constant
field implied by $\chi=ia$; such an imaginary field would
require a constant additional surface term to contribute the
action integral.  Besides this constant term
$\frac{1}{2}R_{0}^{2}$, (9) exactly coincides with the
Lagrangian in (1). For the axionic field case where the symmetry
$\chi=ia$ holds, $\varphi^{2}=-\varphi_{0}^{2}$, with
$\varphi_{0}$ real. Then the solution to the equations of motion
and Hamiltonian constraint is
\begin{equation}
a(\tau)=(m\varphi_{0})^{-1}[(1+2m^{2}R_{0}^{2})^{\frac{1}{2}}
\cosh(2^{\frac{1}{2}}m\varphi_{0}\tau)-1]^{\frac{1}{2}},
\chi=ia(\tau),
\end{equation}
which represents an axionic nonsingular wormhole spacetime.

If we rewrite (9) as a Lagrangian density
$L(\Phi,a)=m_{p}^{2}L(\varphi,a)/a^{4}$, it turns out that,
although the Lagrangian density as written in the form
$L(\Phi,a)$ looks formally similar (except the last term in the
potential) to that for an isotropic and homogeneous Higgs model
in Euclidean time, it however preserves all the symmetries of
the theory intact. Indeed, the field $\Phi$ is not but a simple
imaginary constant $\Phi=i(2\pi^{2}\sigma^{2})^{-\frac{1}{2}}$.
None the less, if $\Phi$ is shifted by some variable $\rho$,
such that $\Phi\rightarrow\phi=i\xi+\rho$, where
$\xi=(2\pi^{2}\sigma^{2})^{-\frac{1}{2}}$, while keeping the
same real scale factor, then symmetry $\chi=ia$, and hence
diffeomorphism invariance, would become broken.  In such a case,
the Lagrangian density $L\equiv L(\phi,a)$ resulting from
$L(\Phi,a)$ is seen (Note that if we let $\Phi$ to be
time-dependent, then the kinetic part of the Lagrangian
$L(\Phi,a)$ becomes
$-[\frac{1}{2}(\frac{\dot{a}}{a^{2}})^{2}\Phi^{2}$
$+(\frac{\dot{a}}{a^{2}})\Phi\frac{\dot{\Phi}}{a}
+\frac{1}{2}(\frac{\dot{\Phi}}{a})^{2}$])
to describe a typical Higgs model in the isotropic and
homogeneous euclidean framework for a charge-$Q$ field $\phi$, a
massless gauge field $A\equiv A(\tau)=\frac{TrK}{eQ}$ (with $e$
an arbitrary gauge coupling and $K$ the second fundamental
form), and variable $tachyonic$ mass $\mu=\frac{1}{a}$, as now
the last term in the potential does not depend on $\phi$ and
becomes thereby harmless for the Higgs model. In principle, the
Lagrangian could have been written in a form other than the
given in (9) by a different use of symmetry $\chi=ia$, allowing
the scalar field to become complex afterwards. The reason why
the system should spontaneously choose breaking the symmetry
from the particular Lagrangian form given by (9) simply is that,
in so doing, it will obtain the most stable vacuum
configuration.

The integration constant $R_{0}^{2}$ in (9) corresponds to the
inclusion of axionic surface terms in the action integral. For a
constant real scalar field, such surface terms should produce a
generally different integration constant $R'_{0}^{2}$ whose sign
is the opposite to that for $R_{0}^{2}$. Therefore, the
solutions to the equations of motion and Hamiltonian constraint
corresponding to a constant real scalar field
$\varphi^{2}=\varphi '_{0}^{2}$ are
\begin{equation}
a_{\pm}(\tau)=(m\varphi
'_{0})^{-1}[1\pm(1-2m^{2}R'_{0}^{2})^{\frac{1}{2}}\cosh(2^{\frac{1}{2}}m\varphi
'_{0}\tau)]^{\frac{1}{2}}.
\end{equation}
Solutions $a_{+}$ and $a_{-}$ lie in the two disconnected,
classically allowed regions for which the potential is negative.
It is only $a_{+}$ which can be thought of as giving the metric
of a wormhole. $a_{-}$ is a singular solution confined to be
$\leq a_{-}(0)$, and hence the isotropic manifold provided with
a metric given by $a_{-}$ is disconnected from the wormhole
manifold and whereby from the two asymptotically flat regions,
for any given allowed values of $R'_{0}^{2}$ and $m^{2}$
($m^{2}R'_{0}^{2}\leq\frac{1}{2}$). The values taken by $a_{-}$
along the time interval (0, $\tau_{0}=(2m^{2}\varphi
'_{0}^{2})^{-\frac{1}{2}}\cosh^{-1}[(1-2m^{2}R'_{0}^{2})^{-\frac{1}{2}}]$)
are nonzero and real, inducing the appearance of a real nonzero
disconnected region inside the wormhole manifold. Hence, the
wormhole manifold will be doubly connected [9].

The real classical solution corresponding to the equilibrium
minimum of the potential for axionic fields is $\bar{a}=0$, and
that for real fields are $\bar{a}_{\pm}=\pm(m\varphi
'_{0})^{-1}$. Therefore, relative to the simply connected inner
topology implied by (10), the doubly-connected topology implied
by (11) represents an actual pitchfork bifurcation.

We can see now why not all values of $\varphi_{0}$ and $\varphi
'_{0}$ are allowed semiclassically.  Although one actually could
also gauge the imaginary part of $\phi$ to any value other than
$\xi$, since the action $I$ depends on $\varphi_{0}^{-2}$ or
$\varphi '_{0}^{-2}$ through solutions (10) and (11), very small
values of the constant fields will make this action very large
and hence the semiclassical probability $e^{-I}$ becomes
vanishingly small. Thus, interpreting $\varphi^{2}$ as a control
parameter it turns out that the bifurcation process cannot be
reached in a continuous, deterministic way. Moreover, purely
real values of $\varphi$ are only compatible with symmetry
breaking if the Higgs mechanism is defined in the unitary gauge,
and this requires specific boundary conditions for the path
integral.

Disregarding the constant term of the potential in (9) to make
this potential vanish at its minima, we obtain the same
situation as in Fig.1, with $\omega\sim 1$ in Planck units. If
it were not for the short intervals where sphalerons are
quantically induced, the contribution of the path integral
$<0\mid e^{-HT/\hbar}\mid 0>$ to the quantum state of the system
would simply be a constant factor. However, if such sphaleron
transtions are taken into account, then the contribution of the
path integral $<0\mid e^{-HT/\hbar}\mid 0>_{sph}$ will introduce
a time-dependent factor like (7), with $\omega\sim 1$, in the
full quantum state. Since time separation between the initial
and final states cannot be known, one should integrate over $T$
to finally obtain for the full quantum state
\begin{equation}
\sum_{j=1}^{\infty}\int_{0}^{\infty}dT\Psi[a,\Psi]\Psi[a',\Psi
']e^{-\frac{1}{2}jK_{sph}T},
\end{equation}
where the $\Psi '$s are the wave functions for the initial and
final states. The full state (12) would give the density matrix
for a nonsimply connected wormhole [10]. If we take for the
small intervals where sphaleron transitions occur a most
probable value of the order the Planck time, then
$K_{sph}\sim(\hbar G)^{-\frac{1}{2}}$.  It follows then that in
the limit where either $\hbar\rightarrow 0$ or $G\rightarrow 0$,
or both, the path integral vanishes. This suggests that the
extra level of quantization introduced by quantum sphalerons can
only appear when quantum-gravity effects are considered, or in
other words, quantum gravity involves an extra level of
quantization which is over and above that is contained in
nongravitational quantum theory.

A caveat is worth mentioning finally. It has been pointed out
[8] that a rotation $t\rightarrow +i\tau$ would imply a
repulsive gravitational regime.  Nevertheless, a probability
functional which is factorizable as a product of equal wave
functions [11] can no longer represent the ground state if
diffeomorphism invariance is preserved, for all eigenenergies
are strictly zero [5]. However, if diffeomorphism invariance is
broken, so that a ground state as (12) becomes well defined,
then such a ground state would be below the barrier for
$t\rightarrow +i\tau$, in a situation where the Euclidean action
is negative and corresponds therefore to a positive
gravitational constant and hence to attractive gravity.

$Acknowledgements$. This work was supported by a $CAICYT$
Research Project N' PB91-0052.

\pagebreak

[1] A.A. Belavin, A.M. Polyakov, A.S. Schwarz and Yu.S. Tyupkin,
Phys. Lett. B59(1975)85.

[2] G.'t Hooft, Phys. Rev. D14(1976)3432.

[3] N.S. Manton, Phys. Rev. D28(1983)2019.

[4] F.R. Klinkhamer and N.S. Manton, Phys. Rev. D30(1984)2212.

[5] A.D. Linde, {\it Inflation and Quantum Cosmology} (Academic
Press, Boston, 1990).

[6] S. Coleman, {\it The Uses of Instantons} in The Whys of
Subnuclear Physics, ed. A. Zichichi (Plenum Press, New York,
1979).

[7] G. Gaeta, Phys. Rep. 189(1990)1.

[8] For a discussion on the sign of Wick rotation in quantum
gravity, see [5] and also the contributions of A.D. Linde and
S.W. Hawking in {\it 300 Years of Gravitation} (Cambridge Univ.
Press, Cambridge, 1987).

[9] P.F. Gonz\'alez-D\'{\i}az, Phys. Rev. D45(1992)499.

[10] P.F. Gonz\'alez-D\'{\i}az, Nucl. Phys. B351(1991)767.

[11] S.W. Hawking, Phys. Rev. D37(1987)904.

\pagebreak

LEGEND FOR FIGURE

Fig.1.- Bifurcation itinerary from $\varphi^{2}<0$ to
$\varphi^{2}>0$ for the potential in Eqn. (1) for $m=0.5$.

\end{document}